# RetinaRegen: A Hybrid Model for Readability and Detail Restoration in Fundus Images

Yuhan Tang[1], Yudian Wang[2], Weizhen Li[1], Ye Yue[3], Chengchang Pan[3*], and Honggang Qi[3*]

[1] Sichuan University
2 International School,Beijing University of Posts and Telecommunications
3 School of Computer Science and Technology, University of Chinese Academy of Sciences
2022141240194@ stu.scu.edu.cn[1]

**Abstract.** Fundus image quality is crucial for diagnosing eye diseases, but real-world conditions often result in blurred or unreadable images, increasing diagnostic uncertainty. To address these challenges, this study proposes *RetinaRegen*, a hybrid model for retinal image restoration that integrates a readability classification model, a Diffusion Model, and a Variational Autoencoder (VAE). Experiments on the SynFundus-1M dataset show that the proposed method achieves a PSNR of 27.4521, an SSIM of 0.9556, and an LPIPS of 0.1911 for the readability labels of the optic disc (RO) region. These results demonstrate superior performance in restoring key regions, offering an effective solution to enhance fundus image quality and support clinical diagnosis.

## 1    Introduction

Traditional medical imaging methods such as CT, MRI, and X-ray have been integral to diagnosis but are often limited by poor resolution or subjective interpretation. AI, particularly deep learning (DL), is revolutionizing diagnostics by enabling automatic interpretation with significantly higher accuracy. Deep learning models, such as convolutional neural networks (CNNs) and their relevance, have demonstrated the ability to identify diseases like diabetic retinopathy [1], breast cancer [2], pulmonary nodules [3] and predict cardiovascular events [4] by learning complex patterns in imaging data. AI models rival human doctors in some tasks, especially when large datasets are used for training [5]. However, challenges remain due to the reliance on high-quality, labeled data and the difficulty of handling ambiguous or corrupted images. These limitations can often hinder the application of AI in clinical settings [6].

Variational Autoencoders (VAE) have shown promise in unsupervised learning, particularly in disentanglement and new sample generation. VAEs can efficiently model complex data distributions and learn latent spaces that enable high-quality data generation [7]. They have been successfully applied to reconstruct low-resolution medical images or handle missing data, which is critical in areas like ophthalmology. Diffusion models, a newer class of generative models, have emerged as highly effective in generating sharp and realistic images. These models work by gradually transforming noise into a target image over multiple steps, offering promising results in image synthesis [8].

---

[1] * Corresponding author.

Ophthalmology often involves interpreting high-resolution images such as Optical Coherence Tomography (OCT), fundus images, or retinal scans. However, noise, motion artifacts, and low-resolution scans can render these images difficult to interpret, impeding early diagnosis and treatment [9]. While recent AI-based methods have improved image quality, the complexity and noise in medical images still pose significant challenges. This study aims to propose an innovative retinal image restoration method to address the challenges existing in current techniques regarding image readability, clarity, and fine detail retention. To achieve this, the study integrates a readability classification model, a diffusion model, and a variational autoencoder (VAE) into a novel image restoration framework, named *RetinaRegen*. First, the readability model filters out unreadable images. Then, ResNet34 with self-attention extracts features from readable images to guide restoration. The diffusion model adds noise and restores image details, while the VAE ensures global consistency and preserves fine details. Finally, the restored image is validated by the readability model. Experiments on the SynFundus-1M dataset [10] show that the proposed method achieves a PSNR of 27.4521, an SSIM of 0.9556, and an LPIPS of 0.1911 for the readability labels of the optic disc (RO) region. Through this innovative approach, *RetinaRegen* aims to improve the accuracy and effectiveness of retinal image restoration, offering new technical support for automated medical image analysis.

## 2     Method

### 2.1     Method Overview

We propose a fundus image restoration method that combines a readability classification model, a diffusion model, and a Variational Autoencoder (VAE). The overall framework extracts features from readable images as conditional information to guide the stepwise restoration of unreadable images. First, a pre-trained InceptionV3-based readability classification model is used to assess the readability of input fundus images, screening unreadable images for restoration. For readable images, global and local features are extracted using a multi-head self-attention mechanism combined with the ResNet34 network and then input as conditional information into the restoration model.

In the restoration process, unreadable images are corrupted with noise using the diffusion model, simulating a stepwise degradation process. At each diffusion step, the noisy images are fused with the extracted conditional features and fed into the VAE encoder to generate high-dimensional feature representations that contain both global context and fine details. Finally, the VAE decoder reconstructs these features into clear, readable images. The restored images are re-evaluated by the readability classification model to verify whether the restoration is successful and meets clinical readability standards. This method effectively addresses unreadability issues in critical regions, such as the optic disc, improving the diagnosis of fundus-related diseases.

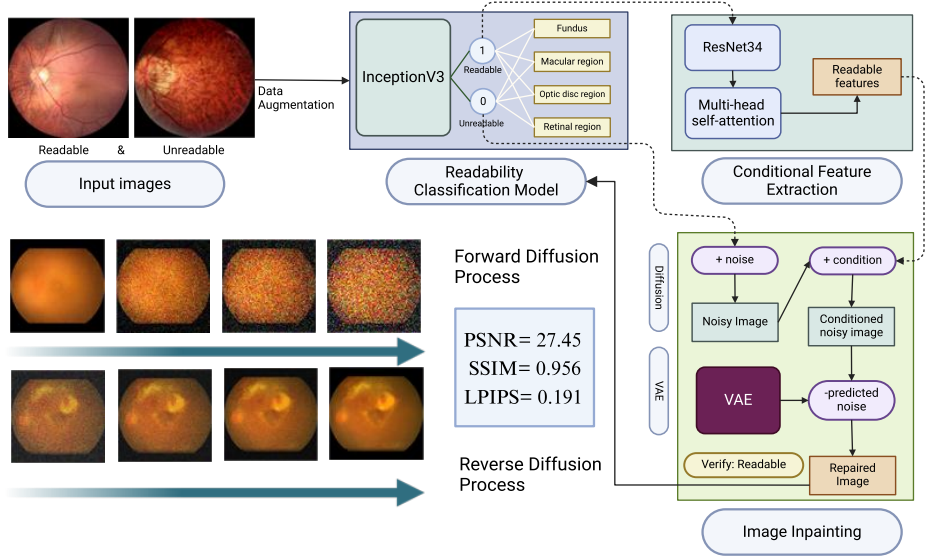

**Fig. 1.** The architecture of the *RetinaRegen*, a hybrid model for retinal image restoration

## 2.2    Readability Classification Model

This study employs a multi-label classification model based on InceptionV3 to evaluate the overall readability of fundus images and the clarity of key regions. The model takes standardized RGB images of size 224×224×3 as input, extracts multi-scale global and local features using residual blocks, and applies global average pooling before outputting four independent binary classification labels. These labels indicate whether the image is a valid fundus image, and whether the macula, optic disc, and retinal regions are readable (see Fig. 1). A Sigmoid activation function is used at the output layer for multi-label classification, with binary cross-entropy as the loss function. The RMSprop optimizer, with a low learning rate (0.0001), is employed to enhance training stability and convergence.

During data preprocessing, input images are normalized and augmented with random horizontal flipping, brightness adjustment, and contrast adjustment to improve robustness and generalization. Samples are divided into training, validation, and test sets in proportions of 64%, 16%, and 20%, respectively. To address class imbalance, class weights are computed based on sample proportions and incorporated into the loss function. The model's performance is evaluated using metrics such as accuracy, precision, recall, and F1-score, along with confusion matrices, ROC curves, and Precision-Recall curves for comprehensive analysis.

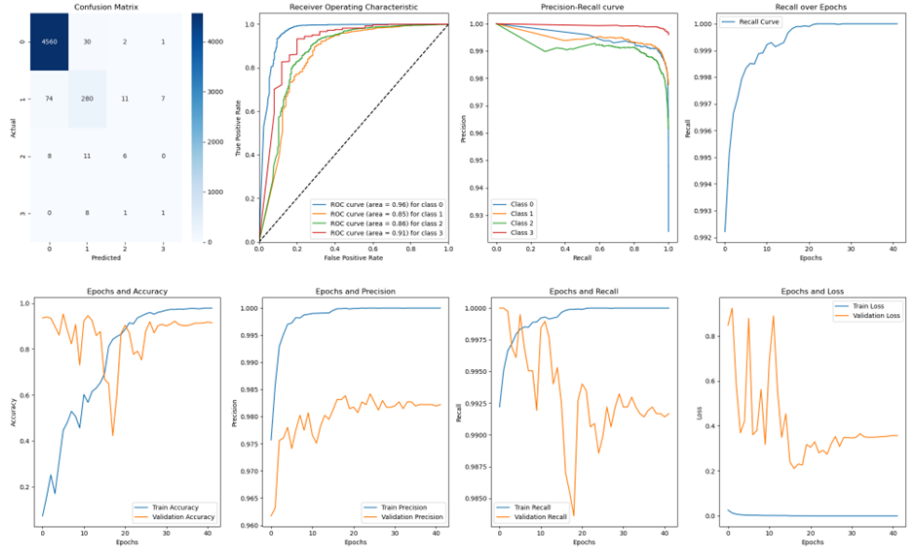

**Fig. 2.** Performance of the readability classification model.

## 2.3    Conditional Feature Extraction

To extract informative features from readable images, this study integrates the ResNet34 architecture with a multi-head self-attention mechanism. Initially, the pre-trained ResNet34 model extracts low-dimensional features from input images. To reduce feature channel complexity, an additional convolutional layer is applied for channel compression, followed by normalization and ReLU activation.

To capture global contextual information, the feature maps are further passed through a multi-head self-attention module with an embedding dimension of 256 and 8 attention heads. By flattening the feature maps and calculating attention weights, the network dynamically focuses on crucial regions, such as the optic disc and its surroundings. The self-attention mechanism enhances the feature maps by retaining key regional details while integrating global features.

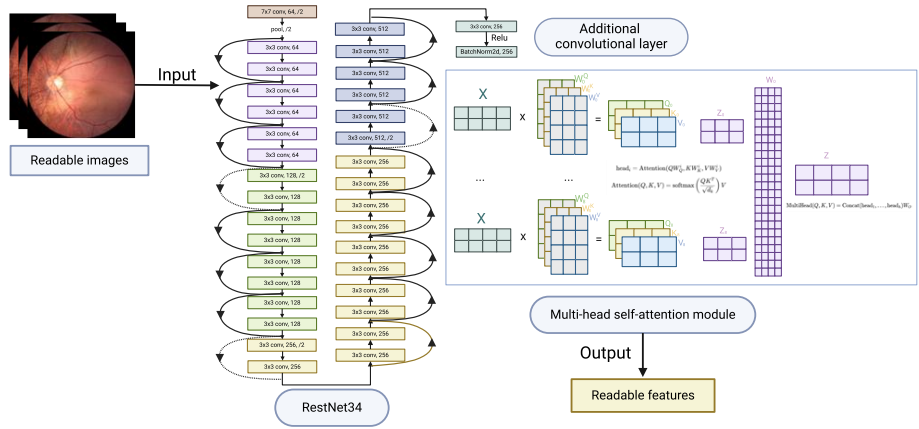

**Fig. 3.** The architecture of the Conditional Feature Extraction

## 2.4 Diffusion Model

The diffusion model gradually adds Gaussian noise to convert the original data distribution into a noise distribution and restores the original data through a reverse diffusion process. During noise addition at time step t, the input image $x_0$ is corrupted as follows:

$$x_t = \sqrt{\overline{\alpha_t}} \cdot x_0 + \sqrt{1 - \overline{\alpha_t}} \cdot \epsilon$$

where $\overline{\alpha}_t$ is the cumulative noise reduction parameter controlling the amount of noise added, and $\epsilon \sim \mathcal{N}(0,1)$ is Gaussian noise. The model generates time-step-specific parameters $\alpha_t$, and $\beta_t$ to ensure consistency between noise addition and reverse diffusion. In the diffusion model, conditional feature information is fused with the diffusion process. Specifically, conditional feature c is concatenated with the diffused image x_t to form the input:

$$x_t^{conditioned} = c + x_t$$

The reverse diffusion process aims to iteratively denoise $x_t^{conditioned}$ and predict $x_0$. By training the model to predict the noise $\epsilon$, restoration is achieved using the formula:

$$x_{t-1}^{conditioned} = \frac{1}{\sqrt{\alpha_t}} \left( x_t^{conditioned} - \frac{1 - \alpha_t}{\sqrt{1 - \overline{\alpha_t}}} \cdot \epsilon_\theta \left( x_t^{conditioned}, t \right) \right) + \sqrt{\beta_t} \cdot \epsilon$$

where $\epsilon_\theta$ is the model-predicted noise, and the mean squared error between $\epsilon_\theta$ and the true noise is used as the training objective. This process enables the gradual restoration of the original data distribution from the noise distribution.

## 2.5 Variational Autoencoder (VAE)

The Variational Autoencoder (VAE) is a generative model that learns the probability distribution of a latent space to enable effective encoding and reconstruction of input data. Its encoder reduces the spatial resolution of input images through three convolutional layers while generating the latent distribution's mean $\mu$ and log-variance $\log \sigma^2$. The reparameterization trick is applied to sample the latent variable z as

$$z = \mu + \epsilon \cdot \sigma$$

where $\epsilon \sim \mathcal{N}(0,1)$, ensuring differentiability for gradient-based optimization.

The decoder reconstructs the original image $\hat{x}$ from $z$. A fully connected layer first maps $z$ back to the convolutional feature space, reshaped as feature maps. These maps are then upsampled through three transposed convolutional layers to restore the original image size. The final layer applies a Sigmoid activation function to constrain pixel values to $[0,1]$, aligning with the input image's value range.

The VAE loss function comprises a reconstruction loss and a regularization term. The reconstruction loss calculated using Binary Cross-Entropy (BCE), measures pixelwise differences between $\hat{x}$ and $x$.

$$\mathcal{L}_{BCE} = \sum_i [x_i \log(\hat{x_i}) + (1 - x_i) \log(1 - \hat{x_i})]$$

The Kullback-Leibler (KL) divergence, ensures the latent distribution $q(z \mid x)$ approximates a standard normal distribution $p(z)$.

$$\mathcal{L}_{\mathcal{KL}} = \frac{1}{2}\sum_j \left(\exp\left(\log \sigma_j^2\right) + \mu_j^2 - 1 - \log \sigma_j^2\right)$$

The combined loss enables the VAE to effectively learn latent representations while preserving input data characteristics.

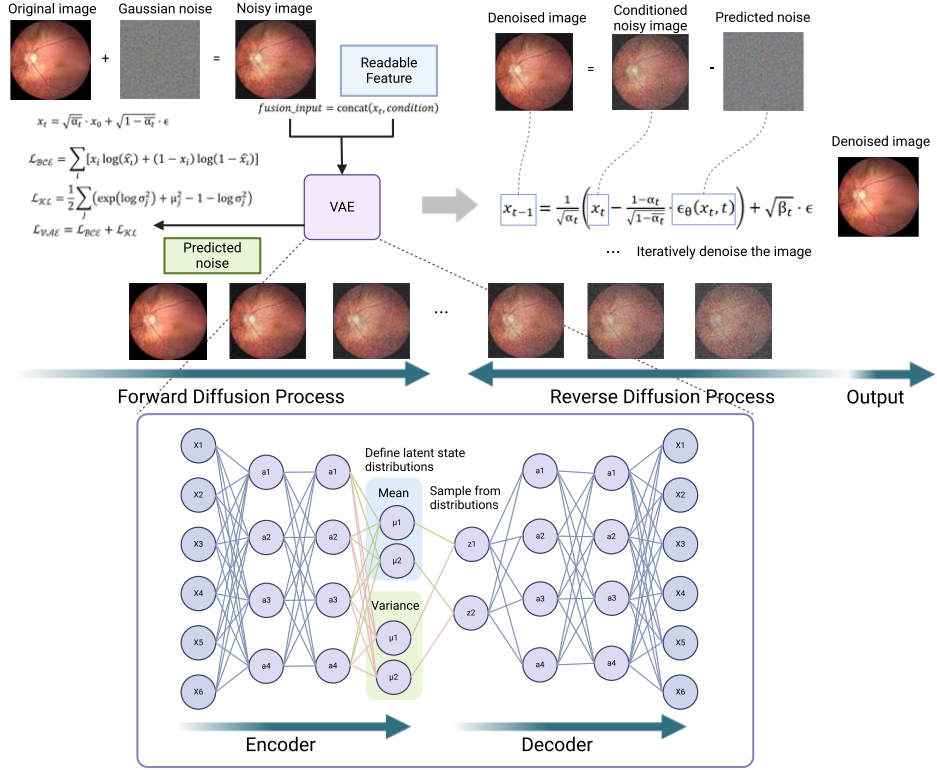

**Fig. 4.** The architecture of the VAE and Diffusion Model

## 3 Experiments

### 3.1 Experimental Setup

The experiments in this study are based on the SynFundus-1M dataset, which includes annotations for 11 types of fundus diseases as well as readability labels for the images. This dataset covers a wide range of real-world image quality issues. The primary task of the experiments is to restore unreadable images, focusing specifically on the optic disc (RO) region. To evaluate the quality of the restored images, this study employs three widely used metrics: Peak Signal-to-Noise Ratio (PSNR), Structural Similarity Index (SSIM), and Learned Perceptual Image Patch Similarity (LPIPS). PSNR reflects the global noise level of the images, SSIM emphasizes structural consistency, and LPIPS evaluates the perceptual similarity between the restored images and the target images. These complementary metrics provide a comprehensive quantitative analysis of the experimental results.

### 3.2 Comparative Experiments

**Comparison of Base Network Architectures.**

To compare the impact of different backbone architectures on the generative performance of the diffusion model, this study evaluated five mainstream architectures: U-Net, U-Net++, ResNet, DenseNet, and VAE, while keeping other modules consistent. ResNet and DenseNet retained the U-Net's encoder-decoder structure but used ResNet's residual blocks and DenseNet's dense blocks, respectively, for feature extraction. The experimental results show that VAE achieved the best performance in terms of detail preservation and structural integrity. The other models, with PSNR values fluctuating between 9.4982 and 10.5970, SSIM values ranging from 0.0773 to 0.2226, and LPIPS between 0.5300 and 0.7857, demonstrated significantly worse performance compared to VAE, which achieved a PSNR of 27.4521, an SSIM of 0.9556, and an LPIPS of 0.1911, demonstrating the strongest generative ability due to its superior latent space learning, which is more effective for maintaining image details and structural integrity.

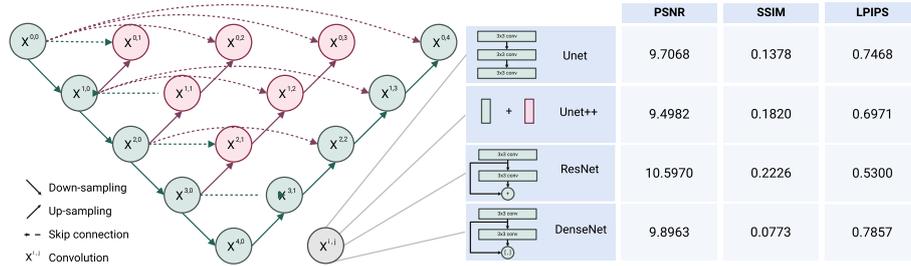

**Fig. 5.** Comparison of Restoration Performance Across Different Base Network Architectures

**Comparison of Readable Image Feature Extraction Methods.**

This experiment investigates the effect of different feature extraction methods on image restoration performance, including self-attention mechanisms and their combinations with various network architectures (Res18, Res34, Res50, AlexNet, MobileNet, ResNeXt, and VGG). The results show that while different feature extraction methods lead to some variation in PSNR, SSIM, and LPIPS, the overall differences in restoration performance are relatively small, with PSNR and SSIM varying between 24.8 and 27.5, and LPIPS ranging from 0.15 to 0.24. Ultimately, the combination of self-attention with Res34 achieved the best performance. Although the performance gap with other methods is not extreme, the overall results indicate that Res34 is the most suitable architecture for this task, effectively optimizing restoration quality while maintaining high detail retention.

**Table 1.** Comparison of Restoration Performance Across Different Feature Extraction Methods

|                          | PSNR    | SSIM   | LPIPS  |
|--------------------------|---------|--------|--------|
| Self-Attention           | 27.0733 | 0.9461 | 0.1736 |
| Self-Attention+AlexNet   | 25.7213 | 0.9445 | 0.2383 |
| Self-Attention+ResNet18  | 25.8058 | 0.9320 | 0.2092 |
| Self-Attention+ResNet34  | 27.4521 | 0.9556 | 0.1911 |
| Self-Attention+ResNet50  | 26.4427 | 0.9446 | 0.1836 |
| Self-Attention+VGG       | 24.8090 | 0.9106 | 0.2038 |
| Self-Attention+ResNeXt   | 26.5226 | 0.9458 | 0.1939 |
| Self-Attention+MobileNet | 26.5648 | 0.9202 | 0.1511 |

**Comparison of Feature Fusion Methods.**

Since we aim to use the extracted readable image features to guide the restoration of unreadable images, this approach encounters issues when merging due to mismatches in resolution and channels. To evaluate the impact of different feature fusion methods on restoration performance, this study compares the following fusion strategies: adding features after upsampling to match resolution and adding features after resolution adjustment using bilinear interpolation, with feature addition either being static (fixed weights) or dynamic (learned weights). The experimental results show that the bilinear interpolation method of adjusting resolution and then adding features statically performs the best. In contrast, the method of adding features after upsampling to match resolution shows poorer performance, with a PSNR of 11.99, SSIM of 0.350, and LPIPS of 0.546. The dynamic addition provides slightly better results than the static addition (PSNR of 12.21, SSIM of 0.681, LPIPS of 0.329), but still cannot match the bilinear interpolation method. The dynamic addition after bilinear interpolation falls short of the static addition, with a PSNR of 15.26, SSIM of 0.713, and LPIPS of 0.271.

## 4      Discussion

The effectiveness and advantages of the proposed method in restoring unreadable images were systematically validated through experiments. Despite the achievements in fundus image restoration, some limitations and areas for improvement remain. First, the experiments in this study are based on readability labels for the optic disc (RO) region from the SynFundus-1M dataset. Thus, the generalizability of the method needs further validation on other image regions and datasets to ensure its applicability in broader clinical scenarios. Second, the readability information used to guide conditional image restoration is generated by pooling features extracted from all readable images, making it static. This limits its ability to provide flexible guidance for individual images. Future studies could explore dynamic conditional information generation and fusion mechanisms, such as adjusting conditional information based on time steps during restoration, to fully utilize the features.

**Acknowledgments.** We thank all affiliates of School of Computer Science and Technology, University of Chinese Academy of Sciences for their valuable feedback. This work was financially supported by Natural Science Foundation of China(grantnumber62271466).

**Disclosure of Interests.** The authors declare that there are no competing intereststoreportinthepaper.